\documentclass[prl,aps,twocolumn,superscriptaddress]{revtex4-2}
\usepackage{graphicx,color,setspace}
\usepackage[utf8]{inputenc}
\usepackage{mathtools}
\usepackage[normalem]{ulem}
\usepackage{tabularx}
\usepackage{enumerate}

\DeclarePairedDelimiter{\norm}{\lVert}{\rVert}

\begin{document}
\title{SAT-assembly: A new approach for designing self-assembling systems}
\author{John Russo}
\affiliation{Dipartimento di Fisica, Sapienza Universit\`{a} di Roma, P.le Aldo Moro 5, 00185 Rome, Italy}
\author{Flavio Romano}
\affiliation{Dipartimento di Scienze Molecolari e Nanosistemi, Universit\`{a} Ca' Foscari di Venezia Campus Scientifico, Edificio Alfa, via Torino 155, 30170 Venezia Mestre, Italy}
\affiliation{European Centre for Living Technology (ECLT) Ca' Bottacin, 3911 Dorsoduro Calle Crosera, 30123 Venice, Italy}
\author{Luk\'{a}\v{s} Kroc}
\affiliation{School of Molecular Sciences and Center for Molecular Design and Biomimetics, The Biodesign Institute, Arizona State University, 1001 South McAllister Avenue, Tempe, Arizona 85281, USA}
\author{Francesco Sciortino}
\affiliation{Dipartimento di Fisica, Sapienza Universit\`{a} di Roma, P.le Aldo Moro 5, 00185 Rome, Italy}
\author{Lorenzo Rovigatti}
\affiliation{Dipartimento di Fisica, Sapienza Universit\`{a} di Roma, P.le Aldo Moro 5, 00185 Rome, Italy}
\author{Petr \v{S}ulc}
\affiliation{School of Molecular Sciences and Center for Molecular Design and Biomimetics, The Biodesign Institute, Arizona State University, 1001 South McAllister Avenue, Tempe, Arizona 85281, USA}

\begin{abstract}
We propose a general framework for solving inverse self-assembly problems, i.e. designing interactions between elementary units such that they assemble spontaneously into a predetermined structure. Our approach uses patchy particles as building blocks, where the different units bind at specific interaction sites (the patches), and we exploit the possibility of having mixtures with several components. The interaction rules between the patches is determined by transforming the combinatorial problem into a Boolean satisfiability problem (SAT) which searches for solutions where all bonds are formed in the target structure. Additional conditions, such as the non-satisfiability of competing structures (e.g. metastable states) can be imposed, allowing to effectively design the assembly path in order to avoid kinetic traps. We demonstrate this approach by designing and numerically simulating a cubic diamond structure from four particle species that assembles without competition from other polymorphs, including the hexagonal structure.
\end{abstract}

\maketitle

\section{Introduction}

Self-assembly defines all the processes by which elementary components organise themselves into ordered structures~\cite{kumar2017nanoparticle}. It occurs ubiquitously in the biological world where proteins, nucleic acids and lipids aggregate with atomistic precision into specific structures that are able to perform a spectacular variety of functions. Nanotechnology has long looked at self-assembly as the most promising avenue for the bottom-up realization of target structures ranging from the nanometer to the micrometer scale as an alternative to top-bottom approaches like micro-patterning and nanolithography.

The goal of the so-called “inverse self-assembly” problem is to design building blocks (or units) that self-assemble, without structural errors, into a desired target structure~\cite{jee2016nanoparticle}. In this process, the interaction between the different units is designed to favour the spontaneous formation of a stable “target” structure. 
The search for the general principles behind self-assembly has attracted several theoretical investigations~\cite{whitelam2015statistical,jacobs2016self}. So far, two promising approaches have emerged: optimization and geometrical approaches.
In optimization algorithms the pair potential is tuned to select a specific target structure. The tuning can be achieved with different strategies such as minimizing the deviation from the target~\cite{rechtsman2005optimized,marcotte2013designeddiamond,zhang2013probing,chen2018inverse}, statistical fluctuations~\cite{miskin2016turning,kumar2019inverse}, and more recently via learning algorithms~\cite{whitelam2021neuroevolutionary,dijkstra2021predictive}.
Geometrical approaches, on the other hand, use geometric features of the target structure to constrain the symmetry of the building blocks~\cite{ducrot2017colloidal,nelson2002toward,manoharan2003dense,zhang2005self,romano2014influence,halverson2013dna,tracey2019programming,park2019design,he2020colloidal,martin2021minimal,mushnoori2021controlling}. While being more readily realizable in experiment, designing the interactions often requires a high degree of geometrical intuition.

Recently, efforts aimed at merging optimization and geometric strategies are beginning to be explored~\cite{patra2017layer,patra2018programmable,chen2018inverse,morphew2018programming,rao2020leveraging,neophytou2021self}. In this manuscript we present one such strategy, called SAT-assembly, where the units are built to match the geometry of the target structure, and whose interactions are optimized by solving a set of Boolean satisfiability (SAT) equations~\cite{romano2020designing}.

Our building blocks are adopted from the family of patchy particle (PP) models, where the assembling units are described by an isotropic repulsion and attractive spots localized on the surface. PP models are used as coarse-grained representations of systems with directional interactions, such as next-generation colloidal particles~\cite{yi2013recent,gong2017patchy,oh2020photo}, proteins~\cite{coluzza2013sequence}, viral capsids~\cite{mosayebi2017beyond,martin2021minimal}, hard-faceted bodies~\cite{rossi2011cubic,smallenburg2012vacancy,van2013entropically}, DNA nanostars~\cite{biffi2015equilibrium,lattuada2020hyperbranched}, double-stranded DNA assemblies, etc.~\cite{zhang2004self,pawar2010fabrication,bianchi2011patchy,romano2011colloidal,bianchi2017limiting}.
Patches represent short-range interacting sites, which can be physically realized with a variety of bonding interactions such as lock-and-key interactions, DNA base pairing, hydrophobic, or dipolar interactions~\cite{pawar2010fabrication}.
As we show in this work, the assembly of PP models is easily translated into SAT problems: the patch type and patch-patch interaction matrix in the PP model are encoded in a Boolean interaction table which is used by SAT to search for solutions over the chosen structure. The advantages of choosing patchy particle models to study self-assembly are numerous: i) the parameters of the model potential (such as the patch width and interaction range) have simple physical interpretations; ii) the thermodynamic behaviour of these models is very well understood~\cite{bianchi2011patchy,de2011phase}; iii) numerous computational techniques have been developed that can considerably accelerate the simulation time required to observe self-assembly phenomena~\cite{rovigatti2018simulate}.

In our approach, the different units bind at specific interaction sites (the patches) that are arranged to match the geometry of the target structure. For example, to self-assemble the cubic diamond structure we consider units whose patches are arranged in a tetrahedral geometry, matching the local environment of the target structure. While in principle one particle with four patches arranged in a tetrahedral geometry would suffice to build up a cubic diamond structure, in reality such a simple choice does not work due to the competition of several polymorphs with comparable free energies~\cite{romano2011crystallization,romano2012patterning,he2020colloidal}. Our approach is designed to solve this problem.

 \begin{figure}[!t]
\begin{center}
\includegraphics[width=7cm]{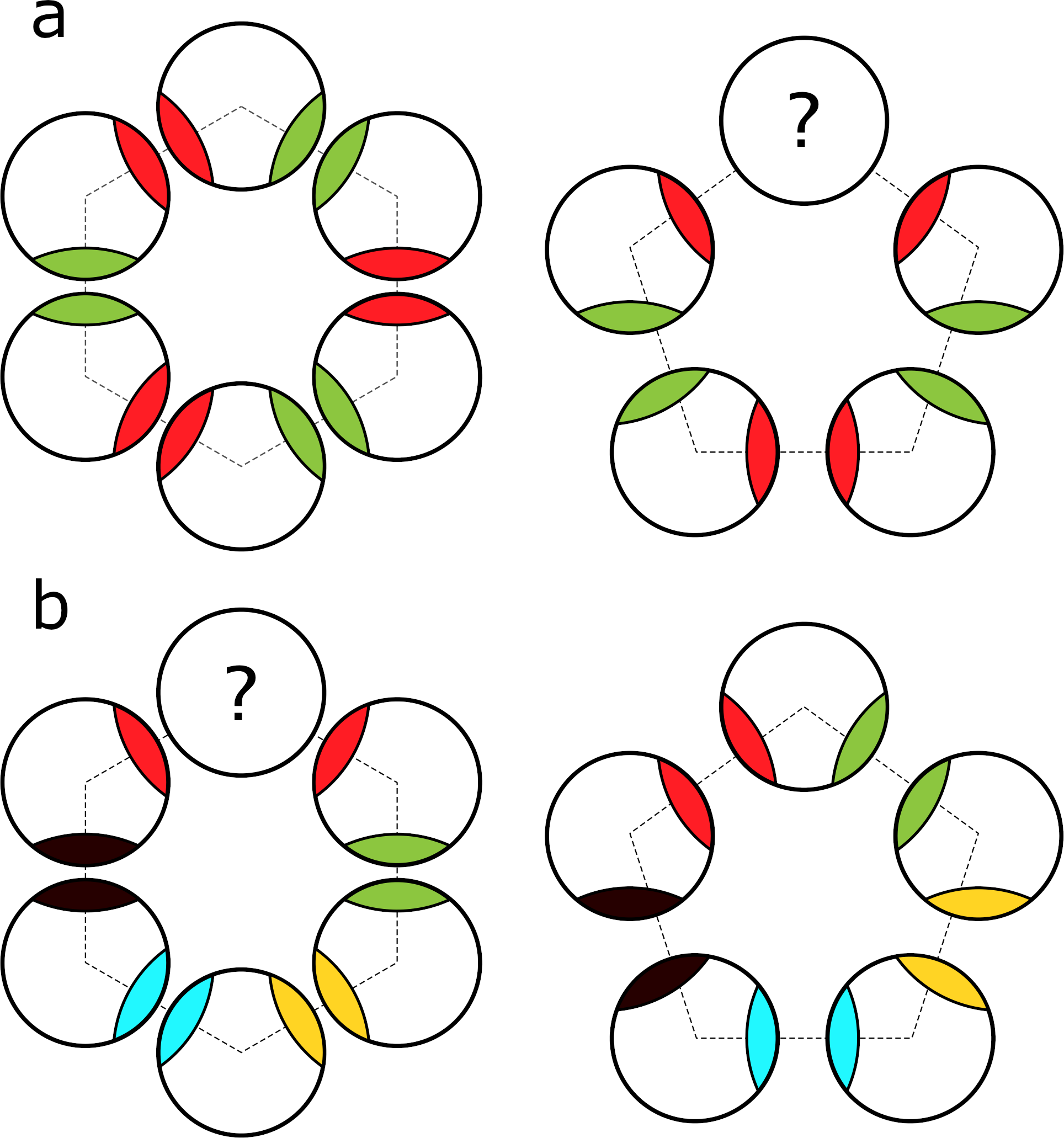}
\caption{Schematic representation of patchy particle designs. \textbf{a.} (1 species, 2 patch types) design allows the formation of six-membered rings but forbids five-membered rings. \textbf{b.} (5 species, 10 patch types) design allows the formation of five-membered rings but forbids six-membered rings. Each patch type interacts with a complementary patch type (as encoded in the interaction matrix), and in the figure each pair is represented with the same color.}
\label{fig:designrules}
\end{center}
\end{figure}

Inspired by biological self-assembly, the patches are designed such that they bind only to specifically selected partner patches (complementary sites). This transforms the inverse self-assembly problem into a “coloring” problem, where each patch has a color assigned to it, and the interaction table between different colors is determined such that all bonds are formed in the target structure.
In Fig.~\ref{fig:designrules} we illustrate the coloring problem for a simple example: the assembly of hexagonal and pentagonal rings. The main difficulty stems from the desire to avoid competing structures that are compatible (either fully or partially) with the interaction table and can appear as kinetic traps during the assembly process as well as competing ordered structures. The study of protocols that help avoiding kinetic traps is an active area of research~\cite{ronceray2017suppression,trubiano2021thermodynamic,bupathy2021temperature}. In our approach, the avoidance of kinetic traps is embedded directly in the interactions. Choosing a small number of colors makes the design easier but usually generates solutions that are compatible with multiple competing structures. On the contrary, specifying a large number of colors makes the search of coloring solutions exponentially harder.

Instead of relying on geometrical intuition to design units that would assemble into a limited set of selected structures~\cite{romano2012patterning,pnasinpress}, our framework solves the coloring problem for any desired structure in a fully automated way. 
The central idea is to convert the problem of “inverse self-assembly design” into a Boolean satisfiability problem (SAT) for the patch colors and the interaction table. All the information on the building blocks is encoded in binary variables and all the rules that specify the bonding geometry are translated into a set of logical clauses. A SAT solver is then used to find solutions to the design problem. The SAT problem is a NP-complete problem, and no polynomial-time solution algorithm is known to exist, but recent advances in the development of SAT solvers have made it possible to solve problems comparable to ours (involving tens of thousands of variables and hundreds of thousand clauses) with relative ease.

In this manuscript, we first briefly introduce SAT, and then describe how to formulate any self-assembly design problem into a SAT problem. We then provide an example application that extends the results we previously obtained in Ref.~\cite{romano2020designing} and use our framework to find a new, previously unreported, patchy particle system that homogeneously nucleates a cubic diamond crystal and consists only of four distinct patchy particle species that are mixed in stoichiometric ratio.  

\section{Boolean Satisfiability Problem}
Boolean satisfiability problems (SAT) are a well-studied topic of interest to computer science and the theory of computational complexity. The general formulation of SAT is to find if there exists a solution in terms of $K$ binary variables $x_i$, where $x_i$ is either 0 (false) or 1 (true), such that a set of clauses $C_j$ are all true at the same time. Each respective clause $C_j$ consists of a subset of variables $x_i$ (or their negations $\neg x_i$) connected by an OR operation ($\lor$ symbol), and all clauses are connected by an AND operation ($\land$ symbol). For example, consider the following SAT problem: Find assignment to three binary variables $x_1$, $x_2$, $x_3$ such that all following clauses are satisfied:
\begin{align}
 C_1 &= x_1 \lor \neg x_2 \\
 C_2 &= \neg x_1 \lor \neg x_3 \lor x_2 \\
 C_3 &= \neg x_1 \lor  x_3 \\
 C_4 &= \neg x_2 \lor \neg x_3 
\end{align}
The problem is to find $x_i$ such that 
\begin{equation}
    C_1 \land C_2 \land C_3 \land C_4 
\end{equation}
is true. One can check that the solution with $x_1 = x_2 = x_3 = 0$ (i.e. all variables set to false) satisfies the constraints, and the problem is hence satisfiable. Obviously, it is also possible to have a task where no solution exists, such as in the following trivial example:
\begin{align}
 C_1 &= x_1 \\
 C_2 &= \neg x_1
\end{align}
where it is impossible to satisfy condition $x_1 \land \neg x_1$. However, for complex tasks with tens of thousands to millions of variables and clauses, it is obviously much more challenging to decide if the problem is satisfiable, and it quickly becomes impossible to combinatorially check all possible values of the variables, as for $K$ variables there are $2^K$ possible states in the solutions space.

SAT problems belongs to the NP-complete complexity class, which includes other famous tasks such as the graph coloring problem or the traveling salesman problem~\cite{NPcomplete}. 
Due to the their importance for the theory of computational complexity as well as for practical discrete optimization tasks, significant effort has been dedicated to the development of efficient SAT solvers. The annual SAT solver competition evaluates the state-of-the-art SAT solving algorithms at a series of tasks. While SAT problems are NP-complete, there are still many instances of SAT problems where solutions can be found very quickly, and modern algorithms have been shown to be able to solve SAT tasks consisting of up to millions of clauses and variables. For the assembly problems presented here, we use MiniSAT~\cite{een2005minisat}, one of the most popular and effective tools for SAT. In our prior work \cite{romano2020designing}, we have also used the WalkSAT and MapleSAT solvers~\cite{liang2018machine,papadimitriou1991selecting,xiao2017maplelrb}. We found that for different target structure design and distinct set of constraints, they sometimes performed differently, with one being able to find a solution (or prove unsatisfiability) within few minutes, while other needing tens of minutes to hours. As opposed to MiniSAT and MapleSAT, WalkSAT cannot be used to prove that a problem is not satisfiable, as it randomly searches the solution space, but we have found that for certain satisfiable problems it performs the fastest. As the choice of optimal solver is problem-dependent, we will be using MiniSAT for the example considered in this work. Overall, for the design problems considered in Ref.~\cite{romano2020designing}, we found %
MapleSAT (which was the top-scoring solver in the SAT 2018 competition \cite{heule2019sat}) 
performed the best, but there were still instances where MiniSAT was faster.

There exists a strong track record of statistical mechanics researchers using the theory of spin glasses to gain insight into the nature of SAT-problems \cite{monasson1999determining,kirkpatrick1994critical,mezard2001bethe,mezard2003two} and identifying regimes in terms of numbers of clauses and variables where SAT problems are ``difficult'' or ``easy'' to solve.
However, the power of SAT solving software has been so far overlooked in the context of self-assembly problems. As we discuss in this paper, it offers a very powerful framework that can complement molecular simulations and effectively find solutions to discrete optimization problems of assigning interactions between particles to achieve self-assembly into desired structures. It is the excellent performance of the available SAT solvers that is the main motivation of our effort to formulate inverse design problem for self-assembly as a SAT problem. As we show below, it is the ability to quickly find solutions and check them against the competing structures with the SAT solvers that enables us to design patchy particles that assemble into any desired structure.

\section{Mapping of The Inverse Design Task to a Boolean Satisfiability Problem}

\begin{figure*}[t]
    \centering
    \includegraphics[width=0.99999\textwidth]{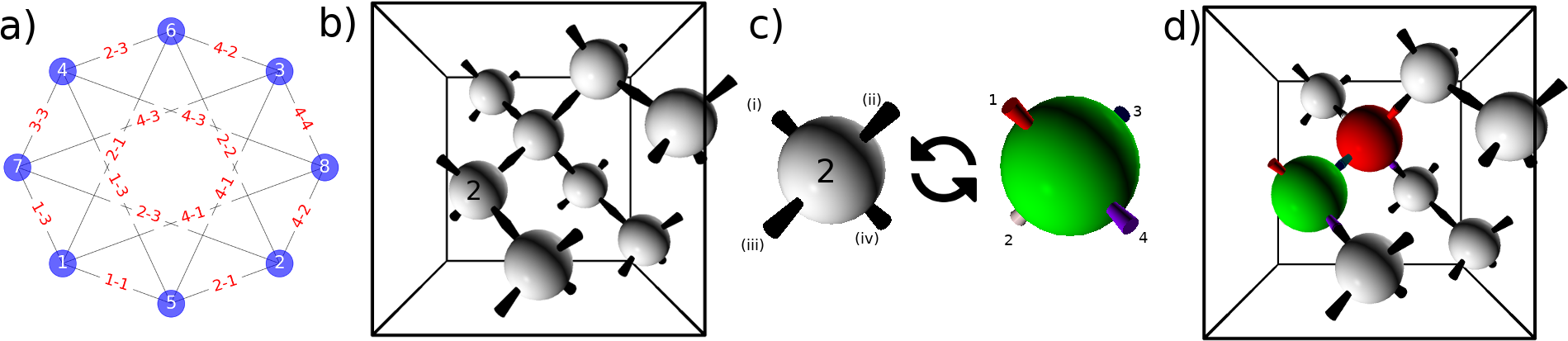}
    \caption{ {\bf a)} A topology representing  the unit cell (with eight positions) of the cubic diamond lattice, showing each lattice position connected to four other positions (interacting slot numbers on each respective positions are shown as link labels). {\bf b)} A schematic 3D representation of the unit cell the of cubic diamond lattice consisting of 8 positions (grey spheres), each bound to its neighbors (using periodic boundary conditions) via numbered ``slots'', shown in black. Each position is assigned a number from 1 to 8, and position number 2 is highlighted as an example. {\bf c)} Position number 2 with its fours slots, numbered from (i) to (iv). The design problem seeks to find patchy particle species with colored slots that are inserted into the positions in the lattice so that their patches (numbered from 1 to 4) overlap with the slots. There are 12 different orientations that a patchy particle with tetrahedral geometry can be positioned into a position so that its patches overlap with respective slots. The twelve possible arrangements are listed in Table \ref{tab:rotation}. {\bf d)} When patchy particles are inserted into the positions, the interacting patches have to be colored with compatible colors. In this schematic example, there are two species, red and green, inserted into two respective positions. The task of the design algorithm is to find particle species with specified patch coloring, color interaction matrix, and patchy particle placements into the respective lattice positions so that all patches are bound to a patch with a compatible color and hence all the bonds in the lattice are satisfied.}
    \label{fig_schematics}
\end{figure*}

In the following we will discuss how to formulate a design problem for patchy particles (PP) as a SAT task.
We focus on designing PP systems which can consist of multiple $N_p$ distinct PP types with fixed geometry of patches that is given by the local contact environment in the desired target structure. In the case of the cubic diamond lattice used in our examples below, we hence consider a tetrahedral arrangement of the patches. The design of the building blocks is thus fixed by the choice of the target structure. Each patch of each PP is assigned a color, and the interactions between PPs are determined by specifying a color interaction matrix, which determines which patches can bind to each other (Fig.~\ref{fig_schematics}). The free parameters of the model are thus: the number of patches and their geometry, the total number of particle species $N_p$, and the number of colors $N_c$.
Keeping $N_p$ and $N_c$ flexible allows to tune these parameters to optimize the assembly, and adapt it to experimental conditions.
We do not impose any torsional restrictions on the patch-patch interaction, but the model can also be extended to account for such restrictions. In our models, complementary interactions have the same strength and we do not allow any misbinding.
As the model system for the description provided below, we use a 8-particle unit cell (Fig.~\ref{fig_schematics}a,b) of a cubic diamond lattice, consisting of tetravalent PP with tetrahedral geometry. The presented approach can however be adapted to any other lattice or even finite-size assemblies.

\subsection{Variable definition}
The first step of mapping a PP design into a Boolean Satisfiability problem is to define a set of binary variables $x_i$ that represent the system and can take either 1 (true) or 0 (false) values.

 We first specify the desired lattice topology (Fig.~\ref{fig_schematics}a), given by the $L$ positions that the particles occupy. For a lattice with an 8-particle unit cell we have $L=8$. Each position in the lattice is assigned $V$ slots, each representative of a bond to a neighboring position in the lattice. For the cubic diamond unit cell in Fig.~\ref{fig_schematics}b, each position in the lattice has $4$ neighbors, so $V=4$. Each of the positions in the lattice has hence four ``slots'' (we use roman numbers to number slots in Fig.~\ref{fig_schematics}c), which are connected to the slots of its neighbors. The lattice topology is encoded by the set of connections between the slots. The goal of the design problem is to find a set of PP and patch colorings so they can be arranged into the target lattice in such a way that the neighboring patches that are in contact in the lattice have complementary colors (i.e., negative potential energy). Note that each PP has the same number of patches as there are number of slots in the unit cell. Satisfying the unit cell with a given available set of PP means that we can populate the target unit cell with patchy particles by placing respective PP into their specific positions in the lattice (Fig.~\ref{fig_schematics}c,d). 
 For a PP with tetrahedral symmetry, there are $N_o=12$ different ways it can be arranged into a specific lattice position so that its patches overlap with the slots of the site, as shown in Fig.~\ref{fig_schematics}c.  We call each possible placement into a specific position an ``orientation $o$'', where $o \in [1,N_o]$. The orientation is given by a list of patch numbers that overlap with the slot numbers for a particular orientation.  For instance, $\phi_2$ corresponds to a mapping from the patches $(1,2,3,4)$ to the slots $(1,4,2,3)$, which means that e.g. $\phi_2(1) = 1$ and therefore
 patch number $1$ will overlap with slot (i), $\phi_2(2) = 4$ and hence patch $2$ will overlap with slot (iv), and so on. All orientations for the tetrahedral system are listed in Table \ref{tab:rotation}.

 \begin{table}
 \centering
 \begin{tabular}{c|c}
 \hline
 Orientation $o$ & Mapping $\phi_o$ \\ \hline
 1 & (1,2,3,4) \\ 
 2 & (1,4,2,3) \\
3 & (1,3,4,2) \\
4 & (2,4,3,1) \\
5 & (2,1,4,3) \\
6 & (2,3,1,4) \\
7 & (4,1,3,2) \\
8 & (4,2,1,3) \\
9 & (4,3,2,1) \\
10 & (3,1,2,4) \\
11 & (3,4,1,2) \\
12 & (3,2,4,1) \\
  \hline
  \end{tabular}
 \caption{\label{tab:rotation} List of orientations $o$ for a PP with a tetrahedral symmetry of patch positions.}
 \end{table}
 As mentioned earlier, we also fix the total  number of particle species $N_p$ and number of colors $N_c$ as parameters of our algorithm.
 We next define the binary variables that describe the entire state space of the design problem:
 \begin{enumerate}
     \item {\bf Color interaction variables:} Binary variables $x^{C}_{c_i,c_j}$ are defined for all combinations of $c_i \leq c_j \in [1,N_c]$, i.e. for all possible pairs of colors, numbered from $1$ to $N_c$. If the variable is true (1) it means that the colors can interact, if it is false then a given pair cannot interact. Please note that our definition allows for self-interaction (if $x^{C}_{c_i,c_i} = 1$ for some color $c_i$). In the case of DNA nanotechnology, such a self-interacting binding is realized e.g.~with a palindromic single-stranded DNA overhang. There are $(N_c)(N_c + 1)/2$ of $x^{C}$ variables.
     \item {\bf Patch coloring variables:} We define variables $x^{pcol}_{p,s,c}$ for all particle types $p \in [1,N_p]$, patch number $s \in [1,V]$ and color $c \in [1,N_c]$. If it is true, it means that particle of type $p$ has its $s$-th patch colored in color $c$. If it is false, then it does not have color $c$. There are $N_p V N_c$ such variables.
     \item {\bf Unit cell placement variables:} We define variable $x^L_{l,p,o}$, which is true if position $l \in [1,L]$ in the target unit cell is occupied by a particle type $p \in [1,N_p]$, which is positioned into the lattice using orientation $o \in [1,N_o]$. There are $N_p L N_o$ of such variables. 
     \item {\bf Auxiliary variables:} We further introduce auxiliary variables that will be later used to help formulate clauses that describe the design problem. We define $x^{A}_{l,s,c}$ which is true if the particle which is positioned in position $l \in [1,L]$ in the lattice is oriented in such a way that the slot $s \in [1,V]$ in the position $l$ is occupied by a patch of color $c \in [1,N_c]$. There are $V L N_c$ of these variables.
 \end{enumerate}

The variables above define the state space of the design problem. To translate the design problem into SAT, we next need to define binary clauses the specify the required relations that these variables need to satisfy. 
 
\subsection{Clause definition}
The variables $x^C_{c_i,c_j}$ are defined for each color combination. However, we need to impose that each color can only be complementary to one other color. If the solution is for example that color $1$ interacts with color $2$, it means that for all $c\neq 1$ or $2$ all variables $x^{C}_{1,c}$ and $x^{C}_{2,c}$ need to be 0. To formulate the requirement that in the solution to our design problem each color $c_i$ can only interact with exactly one other color (including possible self-complementarity), we define clauses $\forall c_i,c_j,c_k \in [1,N_c],  c_i \leq c_j < c_k$:
\begin{equation} 
\label{eq_exactly_one_color}
 C^{\rm int}_{c_i,c_j,c_k} = \neg x^{C}_{c_i,c_j} \lor \neg x^{C}_{c_i,c_k}.
\end{equation}
One can check that if we impose now that color 1 and 2 interact, that is $x^{C}_{1,2} = 1$, the above clauses will now include $0 \lor \neg x^C_{1,c}$ that have to be satisfied for all $c \neq 2$ and clauses $0 \lor \neg x^C_{2,c}$  that have to be true for all $c \neq 1$, thus forcing all variables $x^{C}_{1,c}$ and $x^{C}_{2,c}$ other than $x^{C}_{1,2}$ to be 0.

 The set of clauses introduced in Eq.~\eqref{eq_exactly_one_color} are called \textit{exactly one} clause types and they ensure that from a certain subset of variable class, only one can be true at a time.  We will reuse them for other variable sets. Analogously to Eq.~\ref{eq_exactly_one_color}, we therefore define clauses $\forall s \in [1,V], p \in [1,N_p], c_l,c_k \in [1,N_c], c_l \neq c_k: $  $C^{\rm pcol}_{p,s,c_l,c_k}$ that ensure that patch no.~$s$ of PP type $p$ will be assigned exactly one color only 
\begin{equation} 
\label{eq_exactly_one_patch}
C^{\rm pcol}_{p,s,c_k,c_l} = \neg x^{\rm pcol}_{p,s,c_k} \lor \neg x^{\rm pcol}_{p,s,c_l}.
\end{equation}

To ensure that each lattice position $l$ is occupied by exactly one particle type with one orientation assigned to it, we introduce $\forall l \in [1,L], p_i < p_j \in [1,N_p], o_i < o_j \in [1,N_o]$: 
\begin{equation} 
\label{eq_exactly_one_pos}
  C^L_{l,p_i,o_i,p_j,o_j} = \neg x^L_{l,p_i,o_i} \lor \neg x^L_{l,p_j,o_j} .
\end{equation}

 \begin{table}
 \centering
 \begin{tabular}{c|c|c|c}
 \hline
 Position $l_i$ & Slot $s_i$ & Position $l_j$ & Slot $s_j$\\ \hline
1 & 1 & 5 & 1  \\
1 & 2 & 6 & 1  \\
1 & 3 & 7 & 1  \\
1 & 4 & 8 & 1  \\
2 & 1 & 5 & 2  \\
2 & 2 & 6 & 2  \\
2 & 3 & 7 & 2  \\
2 & 4 & 8 & 2  \\
3 & 1 & 5 & 4  \\
3 & 2 & 6 & 4  \\
3 & 3 & 7 & 4  \\
3 & 4 & 8 & 4  \\
4 & 1 & 5 & 3  \\
4 & 2 & 6 & 3  \\
4 & 3 & 7 & 3  \\
4 & 4 & 8 & 3  \\
  \hline
  \end{tabular}
 \caption{\label{tab:diatable} Cubic diamond 8-unit cell topology (also shown in Fig.~\ref{fig_schematics}a): List of lattice positions $l_i$ and $l_j$ that are neighbors in the unit cell of cubic diamond lattice and their respective slot numbers $s_i$, $s_j$ through which they are bound. The topology is also show in Fig.~\ref{fig_schematics}a.}
 \end{table}

Next, we introduce sets of clauses that impose the topology of the unit cell. We define clauses that enforce for all neighbor positions $l_i$ and  $l_j$ in the lattice that are connected by slots $s_i$ and $s_j$ (shown e.g.~in Table \ref{tab:diatable} and Supplementary Table S1 for cubic diamond lattice with 8 and 16 unit cell respectively) the slots have to be assigned colors $c_i,c_j$ that can bind to each other: $\forall c_i \leq c_j \in [1,N_c]:$
\begin{equation}
\label{eq_slots}
 C^{\rm lint}_{l_i,s_i,l_j,s_j,c_i,c_j} = \neg x^A_{l_i,s_i,c_i} \lor \neg  x^A_{l_j,s_j,c_j} \lor x^C_{c_i,c_j}.
\end{equation}
One can easily verify that the above formulation of such a condition is equivalent to 
$$(x^A_{l_i,s_i,c_i} \land  x^A_{l_j,s_j,c_j}) \implies x^C_{c_i,c_j},$$
which is more intuitively understandable. However, the formulation in Eq.~\eqref{eq_slots} complies with the requirement of SAT, where each individual clause has to be a set of OR statements connecting a set of binary variables or their negation.

Finally, we need to specify that if a lattice position $l$ is occupied by a particle type $p$ with assigned orientation $o$ (in which case $x^L_{l,p,o} = 1$), the slots $s$ of position $l$ are set to have the color of the patch occupying them:  $\forall l \in [1,L], p \in [1,N_p], o \in [1,N_o], s \in [1,V], c \in [1,N_c]$:
$$
  x^L_{l,p,o} \implies \left( x^A_{l, s, c} \iff x^{\rm pcol}_{p, \phi_o(s), c} \right) ,  %
$$
 which can be equivalently rewritten as 
\begin{align}
 C^{\rm LS}_{l,p,o,c,s} &= \left( \neg  x^L_{l,p,o} \lor \neg x^A_{l, s, c} \lor x^{\rm pcol}_{p, \phi_o(s), c} \right) \land \nonumber \\
 & \land  \left(  \neg  x^L_{l,p,o} \lor  x^A_{l, s, c} \lor \neg x^{\rm pcol}_{p, \phi(s, o), c} \right) 
 \label{eq_latice_equiv}.
\end{align}
so that the problem is formulated in terms of logic OR operations on variable subset, joined by AND clauses, as required by SAT. Note that the above clause definition uses the mapping function $\phi_o$ assigned to orientation $o$ as defined in Table \ref{tab:rotation}.

We further define the following additional clauses to ensure that the solution will require that all $N_p$ particle types are present in the desired target lattice, and also that each color is used at least once for coloring patches:
\begin{equation}
\label{eqallp}
 \forall p \in [1,N_p]: C^{\rm all\,p.}_{p} =  \bigvee_{\forall l \in [1,L], o \in [1,N_o] } x^L_{l,p,o} 
\end{equation}
and analogously we also require all colors to be used:
\begin{equation}
\label{eqallc}
 \forall c \in [1,N_c]: C^{\rm all\,c.}_{c} =  \bigvee_{\forall p \in [1,N_p], s \in [1,V] } x^S_{p,s,c}. 
\end{equation}
For example, the clause in Eq.\eqref{eqallp} for $p=1$ requires that there is at least one position $l$ with some orientation $o$ such that $x^L_{l,1,o}$ is true, i.e., PP of type 1 is present in the unit lattice cell in the solution.

Our choice of Boolean variables and conditions has been optimized to allow solvability of typical self-assembly structures in a reasonable computing time.

\subsection{Solving design problem and eliminating competing structures}
\label{sec:solve}
The final SAT formulation is then just a conjunction of all clauses $C$ defined above in Eqs.~\eqref{eq_exactly_one_color} - \eqref{eqallc}. 
If a solution is found in terms of the variables $x_i$, it can be straightforwardly converted into human-readable form by just listing the variables $x^{\rm pcol}_{p,s,c}$ and $x^{C}_{c_i,c_j}$ that are 1, as their subscripts will specify the patch coloring for each PP type and color interaction matrix.

\begin{figure}[t]
    \centering
    \includegraphics[width=0.3\textwidth]{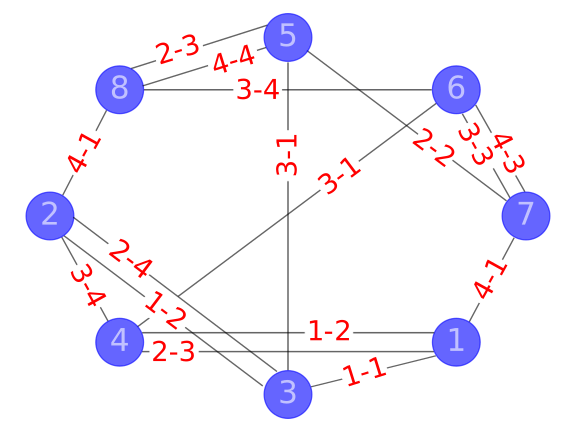}
    \caption{A topology representing the 8 particle unit cell  of hexagonal diamond lattice. Lattice positions are connected by two edges correspond to interaction through periodic boundary condition. \label{fig_hexa}}
   \end{figure}

Note that the SAT solver guarantees that these particles can be arranged into the target lattice, but it is of course still possible that the found solution can also satisfy some other competing lattice.
If competing polymorphs are known beforehand, one can use the SAT formulation to discard solution that can assemble into them. For example, in the case of the cubic diamond lattice, it is already known that hexagonal diamond lattice is an often-encountered competing state, so we harness the SAT formulation to quickly check against this competing lattice. We formulate a new SAT problem, where we now use a new lattice (specified by the topology of the hexagonal diamond lattice, shown in Table~\ref{tab:hexagonal}). We use all clauses from Eqs.~\eqref{eq_exactly_one_color}--\eqref{eq_latice_equiv} (where clauses in Eq.~\eqref{eq_slots} use slots $s_i, s_j$ and lattice positions $l_i, l_j$ as given by the topology of hexagonal unit cell (given in Table \ref{tab:hexagonal} for 8-unit cell and Supplementary Table S2 for 32-unit cell)). We additionally add new clauses that constrain the variables $x^{ \rm pcol}_{p,s,c}$ and 
$x^{C}_{c_i,c_j}$ to be true, where the indices $p, s$, and $c$ are determined by the set of patchy particle types with their assigned coloring that we want to test (i.e., the solution that came out from the SAT problem applied to cubic diamond lattice). The indices in $x^{C}_{c_i,c_j}$ that are set to $1$ encode the given interaction matrix that we are testing. Note that we did not include clause sets from Eq.~\eqref{eqallp}, as we also want to include the possibility that a subset of patchy particle types can assemble into the competing structure. We further do not include clauses from Eq.~\eqref{eqallc}, which is redundant in this case because it is already satisfied by the fact that we set variables $x^{ \rm pcol}_{p,s,c}$ to true for a combination of patch colorings that already satisfies the condition that each color is used for coloring at least one patch.

If indeed the competing structure can be formed, the SAT solver will find a solution in terms of variables $x^L_{l,p,o}$. The indices $l$, $p$ and $o$ of the variables provide the particle type and orientation for each lattice position, allowing to immediately visualize the way in which particles can assemble into the undesired lattice geometry. If that is the case, it means that the original solution that we found in terms of patchy particle types and coloring can also assemble into the other lattice.
Similarly, other competing lattices (if known or identified in simulations) can be checked as well. In that case, we can reformulate the SAT problem for the original desired lattice using different parameters $N_p$ and $N_c$ and see when the solution is able to exclude alternative lattices. These checks are very fast, for lattices of size of tens (or less) of positions they take only fractions of a second when using MiniSAT. Alternatively, if we want to keep $N_p$ and $N_c$ fixed but find other solutions with the same number of particle types and colors, one new clause has to be added, which is a set of OR operators with the negation of the particular set of variables $x^{C}_{c_i,c_j}$ and $x^{\rm pcol}_{p,s,c}$  that describe the solution we already tested and want to exclude:
\begin{equation}
C_{\rm avoid } = \bigvee_{c_i, c_j \in \mathcal{I}} \neg x^{C}_{c_i,c_j}  \bigvee_{p \in [1,N_p], s \in [1,V], c \in \mathcal{C}_{p,s} } \neg x^{\rm pcol}_{p,s,c} 
\end{equation}
where $\mathcal{I}$ is a set of all pairs of interacting colors and $\mathcal{C}_{p,s}$ is the coloring of $s$-th patch of particle type $p$ in the solution that we want to avoid.

 \begin{table}
 \centering
 \begin{tabular}{c|c|c|c}
 \hline
 Position $l_i$ & Slot $s_i$ & Position $l_j$ & Slot $s_j$\\ \hline
1 & 1 & 3 & 1  \\
1 & 2 & 4 & 1  \\
1 & 3 & 4 & 2  \\
1 & 4 & 7 & 1  \\
2 & 1 & 3 & 2  \\
2 & 2 & 3 & 4  \\
2 & 3 & 4 & 4  \\
2 & 4 & 8 & 1  \\
3 & 3 & 5 & 1  \\
4 & 3 & 6 & 1  \\
5 & 2 & 7 & 2  \\
5 & 3 & 8 & 2  \\
5 & 4 & 8 & 4  \\
6 & 2 & 7 & 4  \\
6 & 3 & 7 & 3  \\
6 & 4 & 8 & 3  \\
  \hline
  \end{tabular}
 \caption{\label{tab:hexagonal} Hexagonal 8-unit cell topology: List of lattice positions $l_i$ and $l_j$ that are neighbors in the unit cell of hexagonal diamond lattice and their respective slot numbers $s_i$ and $s_j$ through which they are bound. The corresponding graphical representation is shown in Fig.~\ref{fig_hexa}.}
 \end{table}

Iteratively, one can then find a solution that satisfies the target lattice and avoid competing assemblies.

\section{Application to the design of the Cubic Diamond Lattice}

\subsection{Simulation model}
\label{sec:frenkel}
In this section we will design a PP model that assembles into the diamond cubic lattice and avoids the hexagonal diamond one. To test the results obtained from the SAT solver as described in Sec.~\ref{sec:solve}, we study the assembly using a Monte Carlo (MC) simulation. We use tetravalent patchy particles with tetrahedral patch arrangement (Fig.~\ref{fig_schematics}c). The positions of the patches, in the orthonormal base associated with the patchy particle, are given as
\begin{align*}
\mathbf{p}_1 &=  R \left( \sqrt{8/9}, 0 , -1/3 \right)  \\
 \mathbf{p}_2 &= R \left( -\sqrt{2/9}, \sqrt{2/3}, -1/3  \right)  \\
 \mathbf{p}_3 &= R \left( -\sqrt{2/9}, -\sqrt{2/3}, -1/3 \right)  \\
 \mathbf{p}_4 &= R \left( 0, 0, 1 \right) ,
 \end{align*}
where $R = 0.5 \, \rm{d. u.}$ (distance units) is the radius of the patchy particle represented by a sphere. Each patchy particle is modeled as a hard sphere, with excluded volume interaction between two particles at distance $r$ defined as 
\begin{equation}
   V_{\rm hs}(r) = \begin{cases}
	\infty & \text{if $r <  2 R $},\\
	0 & \text{otherwise}.
	\end{cases} 
\end{equation}
The interaction between a pair of patches $p_i$ and $q_j$ on distinct particles $i$ and $j$ is modeled through the Kern-Frenkel interaction potential \cite{bol1982monte,kern2003fluid}:
\begin{equation}\label{eq_kern_frenkel}
  V_{\rm KF}(r,\theta_p, \theta_q) = \begin{cases}
	-1 & \text{if $r < 2 R + \delta $ and }\\
	   & \text{ $\cos \theta_p > \cos \theta_{\rm max}$ and}  \\ 
	   & \text{ $\cos \theta_q > \cos \theta_{\rm max}$},\\
	0 & \text{otherwise}.
	\end{cases} 
\end{equation}
where $\delta$ and $\cos \theta_{\rm max}$ specify the range and width of the patches. 
Furthermore, we use $\mathbf{r} = \mathbf{r}_{{\rm cm}_q} -\mathbf{r}_{{\rm cm}_p}$, where $r = \norm{\mathbf{r}}$ is the distance between the centers of mass of the patchy particles $p$ and $q$, to define angles
 \begin{align}
  \cos \theta_p &= \frac{\mathbf{r} \cdot \mathbf{p_i}}{ \norm{\mathbf{r}}  \norm{\mathbf{p_i}} } \\
  \cos \theta_q &= \frac{\mathbf{-r} \cdot \mathbf{q_j}}{ \norm{\mathbf{r}}  \norm{\mathbf{q_j}} } 
 \end{align}
where $\mathbf{p_i}$ is the vector from center of mass of particle $p$ towards patch $p_i$, and analogously for patch $q_j$.
The width ($\delta$) and the angular range ($\cos \theta_{\rm max}$) are not part of the SAT definitions. The only requirement on these parameters is that they should allow each patch to form at most one bond at a time (\emph{one bond per patch} condition). Bond range and width should be modeled on the experimental conditions, and optimized independently depending on the relationships between bond geometry and crystallizability~\cite{russo2021physics}.
The simulation model is implemented within the oxDNA simulation package~\cite{rovigatti2015comparison}, which is mostly used for simulations of coarse-grained models of DNA or RNA, oxDNA/oxRNA \cite{doye2013coarse,vsulc2014nucleotide}, but is also a universal simulation package that also implements other models, including the Kern-Frenkel interaction for patchy particle simulations.

\subsection{Set of particles to assemble cubic diamond lattice}

We apply the SAT design framework to find a set of particles that assemble into a cubic diamond lattice.
Our previous solution from Ref.~\cite{romano2020designing} was found for $N_p = 9$ and $N_c = 31$. Aside from the high number of species required, the solution did not have equal concentration of the particle types in the unit lattice, meaning that particle types would have to be mixed in different ratios, which is an added complication for experimental realization.
In this work we have conducted extensive scans of the solution space for different combinations of $N_p$ and $N_c$ and identified a new solution with $N_p = 4$ and $N_c = 12$, with the four species used at 1:1:1:1 ratio in the assembled 16-unit cubic diamond lattice.
The patchy particle types with their coloring and color interaction rules are shown in Table~\ref{table:four}. In the next section we verify this new solution via Monte Carlo simulation with the Kern-Frenkel potential model (as defined in Sec.~\ref{sec:frenkel}).

\begin{table}[]
\centering
\begin{tabular}{|l|llll|}
\hline
PP species & \multicolumn{4}{|c|}{Patch Coloring}\\
\hline
\mbox{1: }  &(a,11) & (b,10) & (c,10) & (d,8) \\
\mbox{2: }  &(a,9) & (b,3) & (c,4) & (d,4) \\
\mbox{3: }  &(a,1) & (b,7) & (c,1) & (d,3) \\
\mbox{4: }  &(a,12) & (b,2) & (c,6) & (d,5) \\
\hline
\multicolumn{5}{|c|}{Color interactions}\\
\hline
\multicolumn{5}{|l|}{(1,10), (2,8), (12,12), (3,4), (5,11), (6,6), (7,9)} \\
\hline
\end{tabular}
\caption{Designed patchy particles for self-assembly into a cubic diamond crystal lattice. It consists of 4 patchy particle types and 12 colors. The patch coloring is in format (patch number, patch color), where patches are labeled from {\it a} to {\it d} (corresponding to patches from 1 to 4 respectively), and colors are numbered from 1 to 12. The color interaction then lists all pairs of interacting colors, where colors 6 and 12 are self-complementary. \label{table:four}}
\label{tab_ts}
\end{table}

\subsection{Assembly simulations}
\label{sec:ass}

\begin{figure*}[t]
    \centering
    \includegraphics[width=0.9\textwidth]{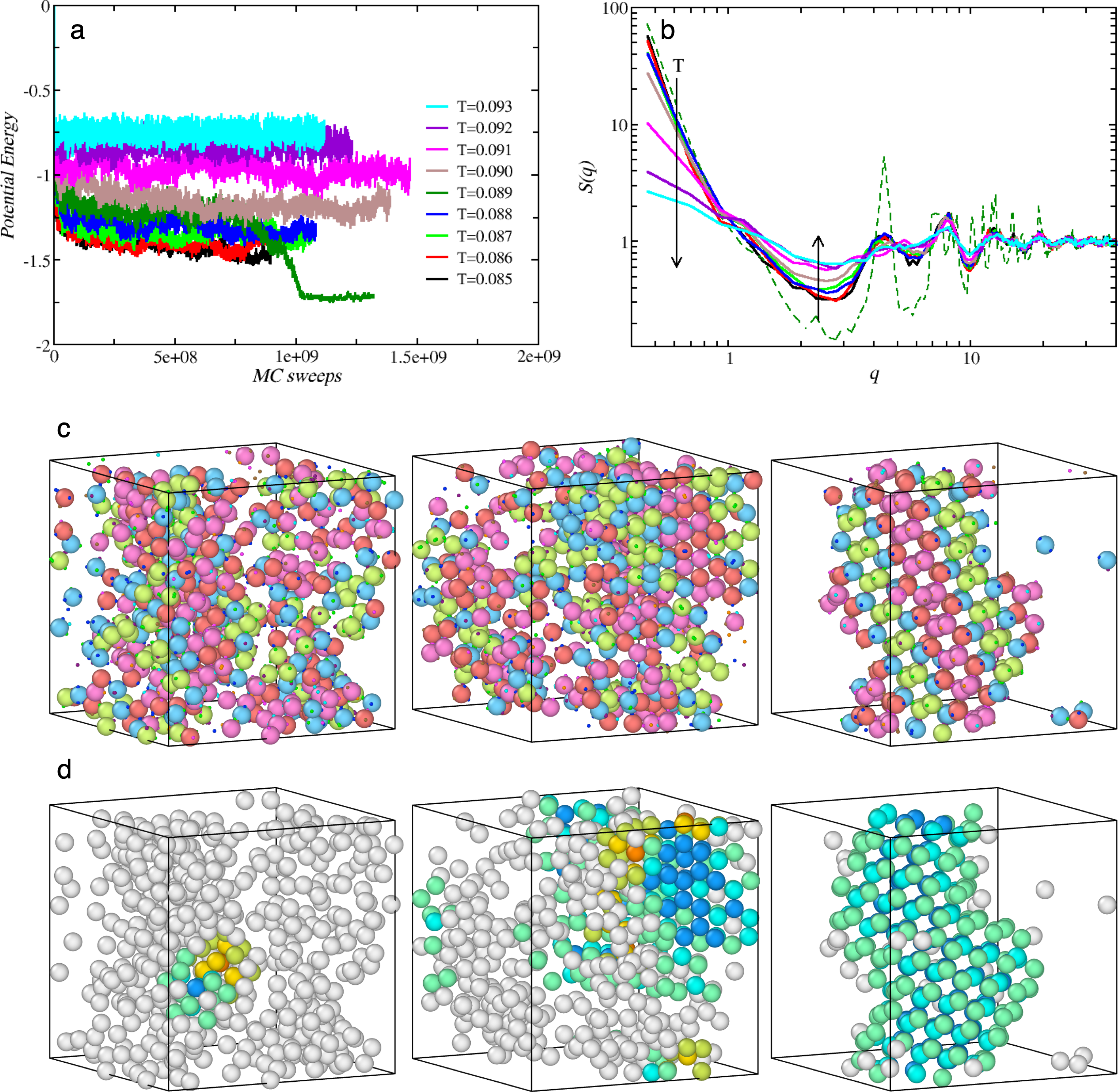}
    \caption{ {\bf a)} Potential energy (in units of $\epsilon$) as a function of MC sweeps for different temperatures (in the legend). As indicated by the energy drop, $T=0.089$ undergoes spontaneous crystallization. {\bf b)} Structure factor ($S(q)$ with the wave-vector in units of $1/2R$) for the same temperatures as in panel a). {\bf c)} Configuration snapshots at MC sweep $7 \times 10^8$, $8.5 \times 10^8$, and $13 \times 10^8$ (from left to right). Particles are colored according to their species (specie 0: red; specie 1: green; specie 2: cyan; specie 3: magenta). Complementary patches have the same color. {\bf d)} Same as in c, but with particles colored according to their phase: liquid (white); cubic diamond (blue); hexagonal diamond (orange). The shade of each color represents the number of crystalline neighbors.  \label{fig_nucleation}}
   
\end{figure*}

We run Monte Carlo simulations of a system of $N=500$ patchy particles (125 particles of each species) colored according to the interactions of Table~\ref{tab_ts}. The Kern-Frenkel potential parameter of Eq.~\ref{eq_kern_frenkel} are the following: $\cos(\theta_\text{max})=0.98$ and $\delta=0.12$. The density is $\rho=0.2$ (constant) and different simulation runs are done at temperatures in the interval $T\in[0.085,0.093]$,  starting from random initial positions and orientations. Fig.~\ref{fig_nucleation}a shows the evolution of the potential energy for all simulation runs. While most temperatures remain in metastable equilibrium for the duration of the simulation, $T=0.089$ displays a clear sign of successful nucleation, i.e. an incubation period followed by a rapid drop of the potential energy.
Fig.~\ref{fig_nucleation}b shows the structure factor of the liquid phase at different temperatures. The strong increase of the signal at $q=0$ with decreasing $T$ shows that all systems are not homogeneous, but have undergone gas-liquid phase separation. The first two peaks of $S(q)$, at $q 2R\approx 4.3$ and $q 2R\approx 8$ respectively, are located in correspondence of  the pre-peak and of the main peak of the diamond crystal structure. This is confirmed by the crystalline peaks for the $T=0.089$ trajectory (dashed line). Fig.~\ref{fig_nucleation}b thus shows that crystallization occurs after the nucleation of a dense liquid phase~\cite{ten1997enhancement}, whose structure is that of a  tetrahedral liquid. 
In Fig.~\ref{fig_nucleation}c and Fig.~\ref{fig_nucleation}d we plot some snapshots of the trajectory at $T=0.089$ during nucleation. In  Fig.~\ref{fig_nucleation}c we color the particles according to their specie, where we can observe the regularity in the specie arrangement in the final crystalline state.
In Fig.~\ref{fig_nucleation}d, particles are colored according to their phase as identified by the Ovito package~\cite{ovito}: liquid (white); cubic diamond (blue); hexagonal diamond (orange).
The snapshots visually confirm that nucleation starts from a de-mixed state and that the final crystalline state is our target structure, i.e. the diamond crystal, without stacking faults.

\section{Conclusion and Outlook}
The search for the general principles behind self-assembly represents a fundamental step towards the promise of nanotechnology to deliver new materials with desired mechanical, optical and thermal properties. The challenges posed by self-assembly are typical of problems with complex free-energy landscapes (e.g. glasses and proteins): given a set of building blocks, there are typically a large number of local free-energy minima that make brute-force approaches computationally intractable both for structure prediction and for its inverse problem, i.e., the design of building blocks that have a desired structure as a global minimum. 
Here we have presented a new framework, named \emph{SAT-assembly}, that adopts two strategies to tame this complexity. The first one is the use of patchy interactions as a general model to encode bonded interactions. The second one is the translation of the coloring problem into a satisfiability problem.
SAT-assembly allows the design of a multicomponent patchy particle system capable of assembling into a target structure while also avoiding competing structures. Our framework allows to avoid both finite-size and long-range ordered competing structures that could interfere with the assembly. Here we have focused on 3D lattice assembly, and provided a detailed explanation of the mapping of the design problem to SAT. Furthermore, we extended our prior work by designing a new (simpler) set of patchy particle types that assembles in the bulk cubic diamond lattice, one of the most sought-after crystal designs in the self-assembly community. We have verified that spontaneous self-assembly is possible with a simulation of the Kern-Frenkel potential, with the target structure being assembled without interference from other competing structures.

Our approach can be also generalized to other periodic lattices, as well as finite-size structures~\cite{boles2016self}. By formulating the design problem as a SAT problem, we can harness the very effective SAT solver tools, which can find solutions to our design problems in the matter of seconds to hours (depending on the size of the target structure and the total number of variables and clauses), and also very quickly check the obtained solution against other competing structures that we want to avoid.

Experimental realization of patchy particle designs is still difficult, but recent advances hold promise that the goal is within reach. 3D DNA nanomaterials, and in particular wireframe DNA origami, that naturally encode sequence complementary, already allow the assembly of structures with the desired interactions~\cite{liu2016diamond,zhang20183d,tian2020ordered,chakraborty2021self}. Other possible realization can be for example via selective patterning of gold nanoparticles \cite{xiong2020three}.

SAT-assembly offers a straightforward and general solution to the problem of inverse self-assembly, which can be employed both to further our fundamental understanding of self-assembling processes and to expand their applications.

\section{Acknowledgments}
JR acknowledges support from the European Research Council Grant DLV-759187. P\v{S} acknowledges support from the ONR Grant N000142012094. JR and P\v{S} acknowledge support from the Universit{\`a} Ca' Foscari for a Visiting Scholarship and from NSF 1931487 - ERC DLV-759187 research collaboration grant.


\section*{References}


\begin{thebibliography}{10}

\bibitem{kumar2017nanoparticle}
S.~K. Kumar, G.~Kumaraswamy, B.~L. Prasad, R.~Bandyopadhyaya, S.~Granick,
  O.~Gang, V.~N. Manoharan, D.~Frenkel, and N.~A. Kotov.
\newblock Nanoparticle assembly: a perspective and some unanswered questions.
\newblock {\em Current Science}, pages 1635--1641, 2017.

\bibitem{jee2016nanoparticle}
A.-Y. Jee, K.~Lou, H.-S. Jang, K.~H. Nagamanasa, and S.~Granick.
\newblock Nanoparticle puzzles and research opportunities that go beyond state
  of the art.
\newblock {\em Faraday discussions}, 186:11--15, 2016.

\bibitem{whitelam2015statistical}
S.~Whitelam and R.~L. Jack.
\newblock The statistical mechanics of dynamic pathways to self-assembly.
\newblock {\em Annual review of physical chemistry}, 66:143--163, 2015.

\bibitem{jacobs2016self}
W.~M. Jacobs and D.~Frenkel.
\newblock Self-assembly of structures with addressable complexity.
\newblock {\em Journal of the American Chemical Society}, 138(8):2457--2467,
  2016.

\bibitem{rechtsman2005optimized}
M.~C. Rechtsman, F.~H. Stillinger, and S.~Torquato.
\newblock Optimized interactions for targeted self-assembly: application to a
  honeycomb lattice.
\newblock {\em Physical review letters}, 95(22):228301, 2005.

\bibitem{marcotte2013designeddiamond}
E.~Marcotte, F.~H. Stillinger, and S.~Torquato.
\newblock Communication: Designed diamond ground state via optimized isotropic
  monotonic pair potentials.
\newblock {\em The Journal of Chemical Physics}, 138(6):061101, 2013.

\bibitem{zhang2013probing}
G.~Zhang, F.~Stillinger, and S.~Torquato.
\newblock Probing the limitations of isotropic pair potentials to produce
  ground-state structural extremes via inverse statistical mechanics.
\newblock {\em Physical Review E}, 88(4):042309, 2013.

\bibitem{chen2018inverse}
D.~Chen, G.~Zhang, and S.~Torquato.
\newblock Inverse design of colloidal crystals via optimized patchy
  interactions.
\newblock {\em The Journal of Physical Chemistry B}, 122(35):8462--8468, 2018.

\bibitem{miskin2016turning}
M.~Z. Miskin, G.~Khaira, J.~J. de~Pablo, and H.~M. Jaeger.
\newblock Turning statistical physics models into materials design engines.
\newblock {\em Proceedings of the National Academy of Sciences}, 113(1):34--39,
  2016.

\bibitem{kumar2019inverse}
R.~Kumar, G.~M. Coli, M.~Dijkstra, and S.~Sastry.
\newblock Inverse design of charged colloidal particle interactions for self
  assembly into specified crystal structures.
\newblock {\em The Journal of chemical physics}, 151(8):084109, 2019.

\bibitem{whitelam2021neuroevolutionary}
S.~Whitelam and I.~Tamblyn.
\newblock Neuroevolutionary learning of particles and protocols for
  self-assembly.
\newblock {\em Physical review letters}, 127(1):018003, 2021.

\bibitem{dijkstra2021predictive}
M.~Dijkstra and E.~Luijten.
\newblock From predictive modelling to machine learning and reverse engineering
  of colloidal self-assembly.
\newblock {\em Nature Materials}, 20(6):762--773, 2021.

\bibitem{ducrot2017colloidal}
{\'E}.~Ducrot, M.~He, G.-R. Yi, and D.~J. Pine.
\newblock Colloidal alloys with preassembled clusters and spheres.
\newblock {\em Nature materials}, 16(6):652--657, 2017.

\bibitem{nelson2002toward}
D.~R. Nelson.
\newblock Toward a tetravalent chemistry of colloids.
\newblock {\em Nano Letters}, 2(10):1125--1129, 2002.

\bibitem{manoharan2003dense}
V.~N. Manoharan, M.~T. Elsesser, and D.~J. Pine.
\newblock Dense packing and symmetry in small clusters of microspheres.
\newblock {\em Science}, 301(5632):483--487, 2003.

\bibitem{zhang2005self}
Z.~Zhang, A.~S. Keys, T.~Chen, and S.~C. Glotzer.
\newblock Self-assembly of patchy particles into diamond structures through
  molecular mimicry.
\newblock {\em Langmuir}, 21(25):11547--11551, 2005.

\bibitem{romano2014influence}
F.~Romano, J.~Russo, and H.~Tanaka.
\newblock Influence of patch-size variability on the crystallization of
  tetrahedral patchy particles.
\newblock {\em Physical review letters}, 113(13):138303, 2014.

\bibitem{halverson2013dna}
J.~D. Halverson and A.~V. Tkachenko.
\newblock DNA-programmed mesoscopic architecture.
\newblock {\em Physical Review E}, 87(6):062310, 2013.

\bibitem{tracey2019programming}
D.~F. Tracey, E.~G. Noya, and J.~P.~K. Doye.
\newblock Programming patchy particles to form complex periodic structures.
\newblock {\em The Journal of Chemical Physics}, 151(22):224506, 2019.

\bibitem{park2019design}
S.~H. Park, H.~Park, K.~Hur, and S.~Lee.
\newblock Design of DNA Origami Diamond Photonic Crystals.
\newblock {\em ACS Applied Bio Materials}, 3(1):747--756, 2019.

\bibitem{he2020colloidal}
M.~He, J.~P. Gales, {\'E}.~Ducrot, Z.~Gong, G.-R. Yi, S.~Sacanna, and D.~J.
  Pine.
\newblock Colloidal diamond.
\newblock {\em Nature}, 585(7826):524--529, 2020.

\bibitem{martin2021minimal}
M.~Mart{\'\i}n-Bravo, J.~M.~G. Llorente, J.~Hern{\'a}ndez-Rojas, and D.~J.
  Wales.
\newblock Minimal Design Principles for Icosahedral Virus Capsids.
\newblock {\em ACS nano}, 2021.

\bibitem{mushnoori2021controlling}
S.~Mushnoori, J.~A. Logan, A.~V. Tkachenko, and M.~Dutt.
\newblock Controlling morphology in hybrid isotropic/patchy particle
  assemblies, 2021.

\bibitem{patra2017layer}
N.~Patra and A.~V. Tkachenko.
\newblock Layer-by-layer assembly of patchy particles as a route to nontrivial
  structures.
\newblock {\em Physical Review E}, 96(2):022601, 2017.

\bibitem{patra2018programmable}
N.~Patra and A.~V. Tkachenko.
\newblock Programmable self-assembly of diamond polymorphs from chromatic
  patchy particles.
\newblock {\em Physical Review E}, 98(3):032611, 2018.

\bibitem{morphew2018programming}
D.~Morphew, J.~Shaw, C.~Avins, and D.~Chakrabarti.
\newblock Programming hierarchical self-assembly of patchy particles into
  colloidal crystals via colloidal molecules.
\newblock {\em ACS nano}, 12(3):2355--2364, 2018.

\bibitem{rao2020leveraging}
A.~B. Rao, J.~Shaw, A.~Neophytou, D.~Morphew, F.~Sciortino, R.~L. Johnston, and
  D.~Chakrabarti.
\newblock Leveraging hierarchical self-assembly pathways for realizing
  colloidal photonic crystals.
\newblock {\em ACS nano}, 14(5):5348--5359, 2020.

\bibitem{neophytou2021self}
A.~Neophytou, V.~N. Manoharan, and D.~Chakrabarti.
\newblock Self-Assembly of Patchy Colloidal Rods into Photonic Crystals Robust
  to Stacking Faults.
\newblock {\em ACS nano}, 15(2):2668--2678, 2021.

\bibitem{romano2020designing}
F.~Romano, J.~Russo, L.~Kroc, and P.~{\v{S}}ulc.
\newblock Designing patchy interactions to self-assemble arbitrary structures.
\newblock {\em Physical Review Letters}, 125(11):118003, 2020.

\bibitem{yi2013recent}
G.-R. Yi, D.~J. Pine, and S.~Sacanna.
\newblock Recent progress on patchy colloids and their self-assembly.
\newblock {\em Journal of Physics: Condensed Matter}, 25(19):193101, 2013.

\bibitem{gong2017patchy}
Z.~Gong, T.~Hueckel, G.-R. Yi, and S.~Sacanna.
\newblock Patchy particles made by colloidal fusion.
\newblock {\em Nature}, 550(7675):234--238, 2017.

\bibitem{oh2020photo}
J.~S. Oh, G.-R. Yi, D.~J. Pine, et~al.
\newblock Photo-printing of faceted DNA patchy particles.
\newblock {\em Proceedings of the National Academy of Sciences},
  117(20):10645--10653, 2020.

\bibitem{coluzza2013sequence}
I.~Coluzza, P.~D. van Oostrum, B.~Capone, E.~Reimhult, and C.~Dellago.
\newblock Sequence controlled self-knotting colloidal patchy polymers.
\newblock {\em Physical review letters}, 110(7):075501, 2013.

\bibitem{mosayebi2017beyond}
M.~Mosayebi, D.~K. Shoemark, J.~M. Fletcher, R.~B. Sessions, N.~Linden, D.~N.
  Woolfson, and T.~B. Liverpool.
\newblock Beyond icosahedral symmetry in packings of proteins in spherical
  shells.
\newblock {\em Proceedings of the National Academy of Sciences},
  114(34):9014--9019, 2017.

\bibitem{rossi2011cubic}
L.~Rossi, S.~Sacanna, W.~T. Irvine, P.~M. Chaikin, D.~J. Pine, and A.~P.
  Philipse.
\newblock Cubic crystals from cubic colloids.
\newblock {\em Soft Matter}, 7(9):4139--4142, 2011.

\bibitem{smallenburg2012vacancy}
F.~Smallenburg, L.~Filion, M.~Marechal, and M.~Dijkstra.
\newblock Vacancy-stabilized crystalline order in hard cubes.
\newblock {\em Proceedings of the National Academy of Sciences},
  109(44):17886--17890, 2012.

\bibitem{van2013entropically}
G.~van Anders, N.~K. Ahmed, R.~Smith, M.~Engel, and S.~C. Glotzer.
\newblock Entropically patchy particles: engineering valence through shape
  entropy.
\newblock {\em Acs Nano}, 8(1):931--940, 2013.

\bibitem{biffi2015equilibrium}
S.~Biffi, R.~Cerbino, G.~Nava, F.~Bomboi, F.~Sciortino, and T.~Bellini.
\newblock Equilibrium gels of low-valence DNA nanostars: a colloidal model for
  strong glass formers.
\newblock {\em Soft Matter}, 11(16):3132--3138, 2015.

\bibitem{lattuada2020hyperbranched}
E.~Lattuada, D.~Caprara, V.~Lamberti, and F.~Sciortino.
\newblock Hyperbranched DNA clusters.
\newblock {\em Nanoscale}, 12(45):23003--23012, 2020.

\bibitem{zhang2004self}
Z.~Zhang and S.~C. Glotzer.
\newblock Self-assembly of patchy particles.
\newblock {\em Nano Letters}, 4(8):1407--1413, 2004.

\bibitem{pawar2010fabrication}
A.~B. Pawar and I.~Kretzschmar.
\newblock Fabrication, assembly, and application of patchy particles.
\newblock {\em Macromolecular rapid communications}, 31(2):150--168, 2010.

\bibitem{bianchi2011patchy}
E.~Bianchi, R.~Blaak, and C.~N. Likos.
\newblock Patchy colloids: state of the art and perspectives.
\newblock {\em Physical Chemistry Chemical Physics}, 13(14):6397--6410, 2011.

\bibitem{romano2011colloidal}
F.~Romano and F.~Sciortino.
\newblock Colloidal self-assembly: patchy from the bottom up.
\newblock {\em Nature materials}, 10(3):171, 2011.

\bibitem{bianchi2017limiting}
E.~Bianchi, B.~Capone, I.~Coluzza, L.~Rovigatti, and P.~D. van Oostrum.
\newblock Limiting the valence: advancements and new perspectives on patchy
  colloids, soft functionalized nanoparticles and biomolecules.
\newblock {\em Physical Chemistry Chemical Physics}, 19(30):19847--19868, 2017.

\bibitem{de2011phase}
D.~de~Las~Heras, J.~M. Tavares, and M.~M.~T. da~Gama.
\newblock Phase diagrams of binary mixtures of patchy colloids with distinct
  numbers of patches: the network fluid regime.
\newblock {\em Soft Matter}, 7(12):5615--5626, 2011.

\bibitem{rovigatti2018simulate}
L.~Rovigatti, J.~Russo, and F.~Romano.
\newblock How to simulate patchy particles.
\newblock {\em The European Physical Journal E}, 41(5):59, 2018.

\bibitem{romano2011crystallization}
F.~Romano, E.~Sanz, and F.~Sciortino.
\newblock Crystallization of tetrahedral patchy particles in silico.
\newblock {\em The Journal of chemical physics}, 134(17):174502, 2011.

\bibitem{romano2012patterning}
F.~Romano and F.~Sciortino.
\newblock Patterning symmetry in the rational design of colloidal crystals.
\newblock {\em Nature communications}, 3:975, 2012.

\bibitem{ronceray2017suppression}
P.~Ronceray and P.~Harrowell.
\newblock Suppression of crystalline fluctuations by competing structures in a
  supercooled liquid.
\newblock {\em Physical Review E}, 96(4):042602, 2017.

\bibitem{trubiano2021thermodynamic}
A.~Trubiano and M.~Holmes-Cerfon.
\newblock Thermodynamic stability versus Kinetic Accessibility: Pareto Fronts
  for Programmable Self-Assembly.
\newblock {\em arXiv preprint arXiv:2104.11341}, 2021.

\bibitem{bupathy2021temperature}
A.~Bupathy, D.~Frenkel, and S.~Sastry.
\newblock Temperature Protocols to Guide Selective Self-Assembly of Competing
  Structures.
\newblock {\em arXiv preprint arXiv:2110.11274}, 2021.

\bibitem{pnasinpress}
A.~Neophytou, D.~Chakrabarti, and F.~Sciortino.
\newblock Facile self-assembly of colloidal diamond from tetrahedral patchy
  particles via ring selection.
\newblock {\em Proceedings of the National Academy of Sciences}, xxx(x):xx--xx,
  2021.

\bibitem{NPcomplete}
{Wikipedia contributors}.
\newblock List of NP-complete problems --- {Wikipedia}{,} The Free
  Encyclopedia, 2021.
\newblock [Online; accessed 1-November-2021].

\bibitem{een2005minisat}
N.~Een.
\newblock MiniSat: A SAT solver with conflict-clause minimization.
\newblock In {\em Proc. SAT-05: 8th Int. Conf. on Theory and Applications of
  Satisfiability Testing}, pages 502--518, 2005.

\bibitem{liang2018machine}
J.~H. Liang, C.~Oh, M.~Mathew, C.~Thomas, C.~Li, and V.~Ganesh.
\newblock Machine learning-based restart policy for CDCL SAT solvers.
\newblock In {\em International Conference on Theory and Applications of
  Satisfiability Testing}, pages 94--110. Springer, 2018.

\bibitem{papadimitriou1991selecting}
C.~H. Papadimitriou.
\newblock On selecting a satisfying truth assignment.
\newblock In {\em FOCS}, volume~91, pages 163--169, 1991.

\bibitem{xiao2017maplelrb}
F.~Xiao, M.~Luo, C.-M. Li, F.~Manya, and Z.~L{\"u}.
\newblock Maplelrb lcm, maple lcm, maple lcm dist, maplelrb lcmoccrestart and
  glucose-3.0+ width in {SAT} competition 2017.
\newblock {\em Proc. of SAT Competition}, pages 22--23, 2017.

\bibitem{heule2019sat}
M.~J. Heule, M.~J{\"a}rvisalo, and M.~Suda.
\newblock SAT competition 2018.
\newblock {\em Journal on Satisfiability, Boolean Modeling and Computation},
  11(1):133--154, 2019.

\bibitem{monasson1999determining}
R.~Monasson, R.~Zecchina, S.~Kirkpatrick, B.~Selman, and L.~Troyansky.
\newblock Determining computational complexity from characteristic ‘phase
  transitions’.
\newblock {\em Nature}, 400(6740):133--137, 1999.

\bibitem{kirkpatrick1994critical}
S.~Kirkpatrick and B.~Selman.
\newblock Critical behavior in the satisfiability of random boolean
  expressions.
\newblock {\em Science}, 264(5163):1297--1301, 1994.

\bibitem{mezard2001bethe}
M.~M{\'e}zard and G.~Parisi.
\newblock The Bethe lattice spin glass revisited.
\newblock {\em The European Physical Journal B-Condensed Matter and Complex
  Systems}, 20(2):217--233, 2001.

\bibitem{mezard2003two}
M.~M{\'e}zard, F.~Ricci-Tersenghi, and R.~Zecchina.
\newblock Two solutions to diluted p-spin models and XORSAT problems.
\newblock {\em Journal of Statistical Physics}, 111(3):505--533, 2003.

\bibitem{bol1982monte}
W.~Bol.
\newblock Monte Carlo simulations of fluid systems of waterlike molecules.
\newblock {\em Molecular Physics}, 45(3):605--616, 1982.

\bibitem{kern2003fluid}
N.~Kern and D.~Frenkel.
\newblock Fluid--fluid coexistence in colloidal systems with short-ranged
  strongly directional attraction.
\newblock {\em The Journal of chemical physics}, 118(21):9882--9889, 2003.

\bibitem{russo2021physics}
J.~Russo, F.~Leoni, F.~Martelli, and F.~Sciortino.
\newblock The physics of Empty Liquids: from Patchy particles to Water.
\newblock {\em Reports on Progress in Physics}, 2021.

\bibitem{rovigatti2015comparison}
L.~Rovigatti, P.~{\v{S}}ulc, I.~Z. Reguly, and F.~Romano.
\newblock A comparison between parallelization approaches in molecular dynamics
  simulations on GPUs.
\newblock {\em Journal of computational chemistry}, 36(1):1--8, 2015.

\bibitem{doye2013coarse}
J.~P. Doye, T.~E. Ouldridge, A.~A. Louis, F.~Romano, P.~{\v{S}}ulc, C.~Matek,
  B.~E. Snodin, L.~Rovigatti, J.~S. Schreck, R.~M. Harrison, et~al.
\newblock Coarse-graining DNA for simulations of DNA nanotechnology.
\newblock {\em Physical Chemistry Chemical Physics}, 15(47):20395--20414, 2013.

\bibitem{vsulc2014nucleotide}
P.~{\v{S}}ulc, F.~Romano, T.~E. Ouldridge, J.~P. Doye, and A.~A. Louis.
\newblock A nucleotide-level coarse-grained model of RNA.
\newblock {\em The Journal of chemical physics}, 140(23):06B614\_1, 2014.

\bibitem{ten1997enhancement}
P.~R. ten Wolde and D.~Frenkel.
\newblock Enhancement of protein crystal nucleation by critical density
  fluctuations.
\newblock {\em Science}, 277(5334):1975--1978, 1997.

\bibitem{ovito}
A.~Stukowski.
\newblock {Visualization and analysis of atomistic simulation data with
  OVITO-the Open Visualization Tool}.
\newblock {\em {MODELLING AND SIMULATION IN MATERIALS SCIENCE AND
  ENGINEERING}}, {18}({1}), {JAN} {2010}.

\bibitem{boles2016self}
M.~A. Boles, M.~Engel, and D.~V. Talapin.
\newblock Self-assembly of colloidal nanocrystals: From intricate structures to
  functional materials.
\newblock {\em Chemical reviews}, 116(18):11220--11289, 2016.

\bibitem{liu2016diamond}
W.~Liu, M.~Tagawa, H.~L. Xin, T.~Wang, H.~Emamy, H.~Li, K.~G. Yager, F.~W.
  Starr, A.~V. Tkachenko, and O.~Gang.
\newblock Diamond family of nanoparticle superlattices.
\newblock {\em Science}, 351(6273):582--586, 2016.

\bibitem{zhang20183d}
T.~Zhang, C.~Hartl, K.~Frank, A.~Heuer-Jungemann, S.~Fischer, P.~C. Nickels,
  B.~Nickel, and T.~Liedl.
\newblock 3D DNA origami crystals.
\newblock {\em Advanced Materials}, 30(28):1800273, 2018.

\bibitem{tian2020ordered}
Y.~Tian, J.~R. Lhermitte, L.~Bai, T.~Vo, H.~L. Xin, H.~Li, R.~Li, M.~Fukuto,
  K.~G. Yager, J.~S. Kahn, et~al.
\newblock Ordered three-dimensional nanomaterials using DNA-prescribed and
  valence-controlled material voxels.
\newblock {\em Nature materials}, 19(7):789--796, 2020.

\bibitem{chakraborty2021self}
I.~Chakraborty, D.~J. Pearce, R.~W. Verweij, S.~C. Matysik, L.~Giomi, and D.~J.
  Kraft.
\newblock Self-assembly dynamics of reconfigurable colloidal molecules.
\newblock {\em arXiv preprint arXiv:2110.04843}, 2021.

\bibitem{xiong2020three}
Y.~Xiong, S.~Yang, Y.~Tian, A.~Michelson, S.~Xiang, H.~Xin, and O.~Gang.
\newblock Three-Dimensional Patterning of Nanoparticles by Molecular Stamping.
\newblock {\em ACS nano}, 14(6):6823--6833, 2020.

\end{thebibliography}
\end{document}


\setcounter{figure}{0}
 \makeatletter 
 \renewcommand{\thefigure}{S\@arabic\c@figure}
 \setcounter{equation}{0}
 \renewcommand{\theequation}{S\@arabic\c@equation}
 \setcounter{table}{0}
 \renewcommand{\thetable}{S\@arabic\c@table}
  \setcounter{section}{0}
 \renewcommand{\thesection}{S\@Roman\c@section}

 \pagenumbering{gobble} 
\maketitle
We include here the topology for a unit cell of cubic diamond crystal lattice 16-unit cell, and for hexagonal diamond lattice 32-unit cell. These cells are created by merging smaller unit cells of cubic diamond and hexagonal diamond respectively.
 \begin{table}
 \centering
 \begin{tabular}{c|c|c|c}
 \hline
 Position $l_i$ & Slot $s_i$ & Position $l_j$ & Slot $s_j$\\ \hline
2 & 4 & 15 & 4  \\
10 & 2 & 14 & 4  \\
4 & 1 & 8 & 3  \\
12 & 3 & 13 & 2  \\
3 & 2 & 6 & 3  \\
1 & 4 & 5 & 4  \\
11 & 4 & 16 & 4  \\
2 & 3 & 5 & 3  \\
4 & 4 & 7 & 2  \\
7 & 4 & 10 & 4  \\
12 & 2 & 14 & 2  \\
3 & 3 & 7 & 3  \\
11 & 1 & 13 & 1  \\
2 & 2 & 6 & 4  \\
4 & 3 & 5 & 2  \\
1 & 1 & 16 & 1  \\
9 & 3 & 14 & 1  \\
8 & 2 & 10 & 1  \\
7 & 1 & 9 & 2  \\
12 & 1 & 16 & 3  \\
3 & 4 & 8 & 4  \\
11 & 2 & 14 & 3  \\
2 & 1 & 16 & 2  \\
1 & 2 & 15 & 1  \\
9 & 4 & 13 & 4  \\
8 & 1 & 9 & 1  \\
10 & 3 & 13 & 3  \\
4 & 2 & 6 & 2  \\
12 & 4 & 15 & 2  \\
3 & 1 & 5 & 1  \\
1 & 3 & 6 & 1  \\
11 & 3 & 15 & 3  \\
  \hline
  \end{tabular}
 \caption{\label{tab:doublediamond} Cubic diamond 16-unit cell topology: List of lattice positions $l_i$ and $l_j$ that are neighbors in the unit cell of the lattice and their respective slot numbers $s_i$, $s_j$ through which they are bound. The unit cell of size 16 is obtained by pasting together two unit cells of size 8.}
 \end{table}

\begin{figure*}
\begin{center}
\includegraphics{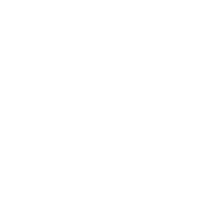}
\end{center}
\end{figure*}

 \begin{table}
 \small
 \centering
 \begin{tabular}{c|c|c|c}
 \hline
 Position $l_i$ & Slot $s_i$ & Position $l_j$ & Slot $s_j$\\ \hline
8 & 4 & 17 & 1  \\
2 & 4 & 23 & 1  \\
10 & 2 & 14 & 2  \\
4 & 1 & 6 & 2  \\
12 & 3 & 29 & 1  \\
22 & 3 & 26 & 2  \\
17 & 3 & 29 & 2  \\
3 & 2 & 7 & 2  \\
27 & 3 & 30 & 4  \\
7 & 3 & 11 & 1  \\
18 & 3 & 30 & 2  \\
1 & 4 & 24 & 1  \\
11 & 4 & 15 & 4  \\
28 & 3 & 29 & 4  \\
8 & 3 & 12 & 1  \\
2 & 3 & 14 & 1  \\
13 & 3 & 28 & 1  \\
23 & 3 & 27 & 2  \\
4 & 4 & 22 & 1  \\
9 & 2 & 13 & 2  \\
17 & 4 & 24 & 2  \\
27 & 4 & 31 & 4  \\
24 & 3 & 28 & 2  \\
7 & 4 & 18 & 1  \\
12 & 2 & 13 & 4  \\
19 & 2 & 21 & 4  \\
26 & 4 & 32 & 3  \\
3 & 3 & 15 & 1  \\
2 & 2 & 7 & 1  \\
20 & 4 & 24 & 4  \\
4 & 3 & 16 & 1  \\
1 & 1 & 5 & 1  \\
14 & 3 & 27 & 1  \\
9 & 3 & 31 & 1  \\
19 & 3 & 31 & 2  \\
10 & 4 & 16 & 2  \\
18 & 2 & 22 & 2  \\
20 & 3 & 32 & 2  \\
3 & 4 & 21 & 1  \\
25 & 3 & 29 & 3  \\
11 & 2 & 14 & 4  \\
5 & 3 & 9 & 1  \\
2 & 1 & 6 & 1  \\
15 & 3 & 25 & 1  \\
6 & 4 & 20 & 1  \\
1 & 2 & 8 & 1  \\
9 & 4 & 15 & 2  \\
19 & 4 & 23 & 4  \\
16 & 3 & 26 & 1  \\
10 & 3 & 32 & 1  \\
21 & 3 & 25 & 2  \\
4 & 2 & 8 & 2  \\
12 & 4 & 16 & 4  \\
28 & 4 & 32 & 4  \\
17 & 2 & 21 & 2  \\
3 & 1 & 5 & 2  \\
25 & 4 & 31 & 3  \\
18 & 4 & 23 & 2  \\
5 & 4 & 19 & 1  \\
26 & 3 & 30 & 3  \\
20 & 2 & 22 & 4  \\
6 & 3 & 10 & 1  \\
1 & 3 & 13 & 1  \\
11 & 3 & 30 & 1  \\
  \hline
  \end{tabular}
 \caption{\label{tab:megahexagonal} Hexagonal 32-unit cell topology, obtained by joining 4 unit cell lattices. List of lattice positions $l_i$ and $l_j$ that are neighbors in the unit cell and their respective slot numbers $s_i$, $s_j$ through which they are bound }
 \end{table}